# Entropical Analysis of an Opinion Formation Model Presenting a Spontaneous Third Position Emergence

*Marcos E. Gaudiano[a,c] , Jorge A. Revelli[b,c]*

[a] *Centro de Investigación y Estudios de Matemática. Consejo Nacional de Investigaciones Científicas y Técnicas.*
[b] *Instituto de Física Enrique Gaviola. Consejo Nacional de Investigaciones Científicas y Técnicas.*
[c] *Universidad Nacional de Córdoba. Facultad de Matemática, Astronomía, Física y Computación.*
*Av. Medina Allende s/n , Ciudad Universitaria, X5000HUA Córdoba, Argentina.*

**Abstract**

Characterization of complexity within the sociological interpretation has resulted in a large number of notions, which are relevant in different situations. From the statistical mechanics point of view, these notions resemble entropy.
In a recent work, intriguing non-monotonous properties were observed in an opinion dynamics Sznajd model. These properties were found to be consequences of the hierarchical organization assumed for the system, though their nature remained unexplained. In the present work we bring an unified entropical framework that provides a deeper understanding of those system features.
By perfoming numerical simulations, the system track probabilistic dependence on the initial structures is quantified in terms of entropy. Several entropical regimes are unveiled. The myriad of possible system outputs is enhanced within a maximum impredictability regime. A mutual structural weakness of the initial parties could be associated to this regime, fostering the emergence of a third position.

## 1. Introduction

Statistical mechanics of disordered systems is an apropriated field for the description of complex systems. This area has found applications in many interdisciplinary areas which can be related to biology, chemistry, economics, social sciences and so on [3, 4, 5].

In particular, the high-throughput methods available for describing social dynamics have drastically increased our ability to gather comprehensive social level information on an ever growing number of situations [5, 6, 7, 8, 9, 10]. These results show that such systems can be thought as a dense network of nonlinear interactions among its components [11], and that this interconnectedness is responsible for their efficiency and adaptability. At the same time, however, such interconnectedness poses significant challenges to researchers trying to interpret empirical data and to extract the deep social principles that will enable us to build new theories and to make new predictions [12].

A central fact, which can be ubiquitously found in nature, is that systems are often organized forming hierarchical structures [13]. Indeed, this organization may play a determining role in system dynamics [1]. Hierachical structure is usually related to a fractal dimension [2, 14, 15]. However, given a set of structures and their interactions, there is not, on one hand, an objective manner to assess whether such hierarchical organization does indeed exist, or to objectively identify the different levels in the hierarchy. On the other hand, by supposing the new structure has already been detected, maybe there is not an explicit natural way to properly describe what is the underlying process behind this emergence.

Entropy is an example of a quantity used in an extensive number of applications [16]. In physics, entropy was shown to be suitable for description of systems in and out of equilibrium. In certain situations, entropy is a measure of a system's evolution in time [17].

Recently, an entropical framework to study a general class of complex sytems was proposed in [2]. The article provides a natural way for characterizing evolution regimes starting from an entropy function associated to hierarchical system configurations. Regimes having maximum incontrollability can be defined and have been already observed in real and different contexts [18, 19].

Besides, in [1] it was shown how a simple opinion formation model develops statistically very different responses if initial conditions are supposed to be random or hierarchically organized. While random initial conditions produce expected results, mainly defined by the relative sizes between the parties, distinct and novel results take place when initial conditions with fractal dimension are assumed. In this case non-monotonous behaviors are obtained. However, despite of the novel properties of the model, no explanation for the source of non-monotonicity was given. In the present work, we show how the general framework of [2] could provide a way to understand not only the source of the non-monotonity underlying the system, but also to explain several other system's properties and behaviors that were so far taken for granted.

The paper is organized as follows: In Section 2 we briefly describe the model studied in [1]. In Section 3 we perform the entropical analysis of the system. Under the framework described in [2], we address motivating discussions about relative entropy productions and the connexion with party winning probabilities. Finally, general conclusions and further extensions of this work are described in Section 4.

## 2. Third position emergence from structured initial conditions

### 2.1. Ideological map

Continuing the ideas explored in [1], a comunity debate is modelled as the time evolution of a squared matrix $M$. Every entry $M_{ij} = \pm 1$ represents the ideological preference of individuals participating actively in the dynamics, while $M_{ij} = 0$ corresponds to persons that neither envolve into debates nor ultimately determine the final opinion state of the system. As in [1], we remark that ideas may be connected to individuals or group of people, in a way that it is possible to image the whole system's dynamics as an idea contend. This makes M to be visualized as an ideological map.

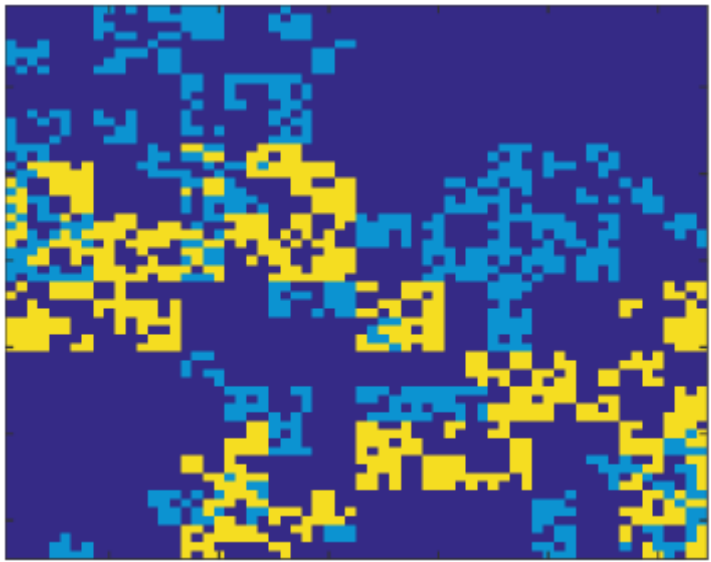

*Figure 1: An example of an ideological map as the ones studied in
[1]. This is a pattern composed of yellow (Y) and light blue (L)
pixels representing the active agents, while the rest (i.e. the dark
blue ones) are associated to the apathic polytical fraction (which
does not participate of the system's debate).*

On Fig.(1) we show an example. We name party L (Y) to the set of light blue (yellow) pixels. The dark blue pixels represents the apathy. In addition, it is worth remarking here that Y and L pixels are not randomly distributed. Specifically, they initially adopt hierarchically organized structures, which are characterized by fractal dimensions $D_Y$ and $D_L$, respectively [1, 15].

The meaning associated to fractal dimensions D will be a kind of organization level of the followers of a particular (Y or L) ideological position [1]. The more organized the party, the higher D value. Higher D means that supporters of an idea conform a more compact pattern, in which an important number of highly convinced people can be found. For low dimensionality patterns, it happens the opposite.

The initial fractions of supporters associated to each party will be equal between each other and denoted by r ($0 \leq r \leq 1$). As in [1], this is done in order to remark the model dependence with respect to the fractal dimension[1].
In addition, the dimension D of every pattern has to run within the following range:

$$J_r = \{D : D_{min} \leq D \leq D_{max}\} \qquad (1)$$

where Dmin and Dmax are monotonous increasing function of r defined in [1] (see also [2]). Thus, the level of participation (or parties' initial sizes) determine the possible organizational structures of the parties.
The patterns are generated without any further constraints, except for the system size which is defined by M side length λ. Throughout this work, it is asumed that λ = 64 for computational simplicity[2].

*2.2. Interaction dynamics*

We consider, as in [1], a slightly modified Sznajd model [10, 20] for the iteration among the ideas[3].
At time t+ Δt just one active site (i, j) is randomly chosen for update[4]. If Mij is the no-neutral position associated to the site, it may change according to

$$M_{ij}(t + \Delta t) = M_{\bar{i}\bar{j}}(t) \qquad (2)$$

where $M_{\bar{i}\bar{j}}(t) \neq 0$ 0 and (ī, j̄) is a second neighbor of (i, j) in the ideological map. Especifically, from the first non neutral neighbors set (simply measured by the Euclidean distance), it is chosen the element (n, m) randomly. This element will be surrounded by non neutral neighbors too, from which it is chosen a third element (ī, j̄) randomly (it is not (i, j)). Certainly, this is a Sznajd-type interaction since the new position adopted at site (i, j) is not related to its first neighbor, but to the position of a far away neighbor, satisfying:

$$(i - \bar{i})^2 + (j - \bar{j})^2 \geq (n - \bar{i})^2 + (m - \bar{j})^2 . \qquad (3)$$

*2.3. Third position emergence*

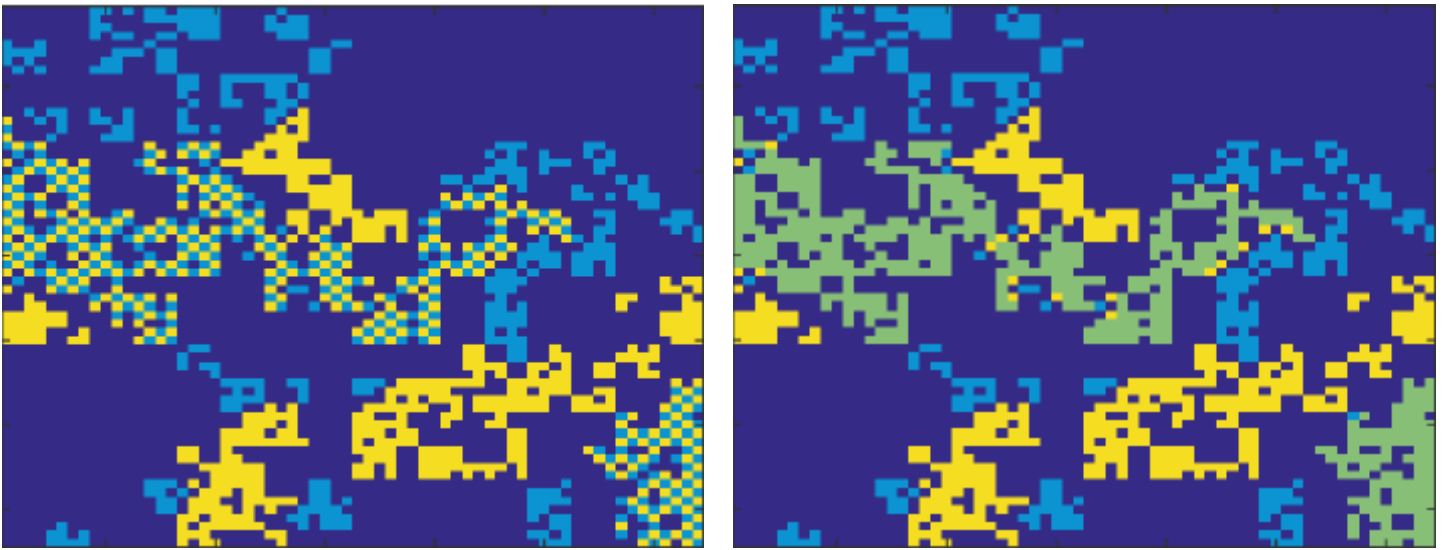

*Figure 2: Left: a possible ideological final pattern assuming Fig.(1)'s initial configuration. Right: the chess-board part of the final state of above is coloured in green (emergent party G).*

---

1 Different levels of participation $r_L$ and $r_Y$ will be studied in [22].
2 for further details about the generation of the patterns see [1]
3 despite being Sznajd's model widely known, the aim of [1] was to be minimalistic and such model was assumed because it has isotropic interactions.
4 we model a system that is not externally driven or forced, so it is plausible to assume an asynchronic update [21].

The system evolves under the above mentioned dynamics. On Fig.(2 , left) it is shown an example of a stationary ultimate state, starting from the initial pattern of Fig.(1). Another stable chess-board-like pattern arises, which is considered as a third emergent party. This new party, composed by a regular mixture of Y and L supporters, is named G. On Fig.(2, right) G supporters have been green colored[5].

## 3. Entropical Analysis

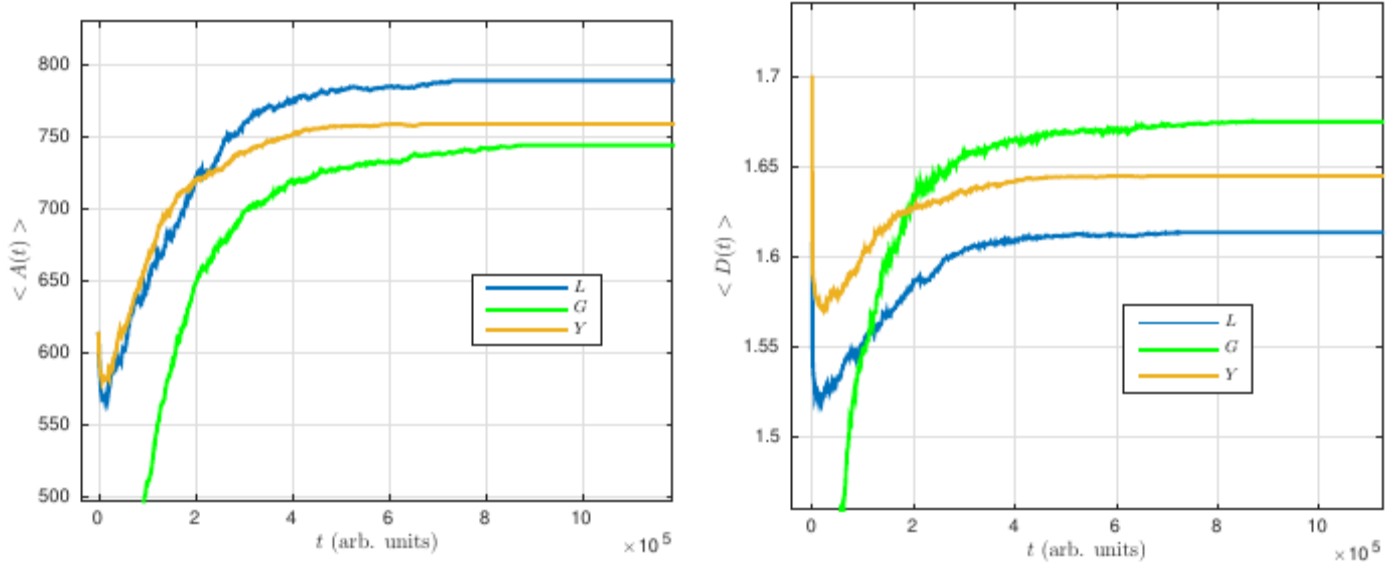

*Figure 3: Average population (left) of followers and dimension (right) of the parties, everytime that either Y , L or G finally wins (Eq.(4)). The total number n of realizations is assumed to be 1000. The initial conditions are: $D_Y$ = 1.70 y $D_L$ = 1.60 and r = 0.15. Every winning party increases its area. When party Y turns out to be the winner, it always reduces its dimensionality (on average).*

In order to develop our ideas, we start by considering the particular case shown in Fig.(3). As mentioned above, Y and L are supossed to have initially each one the same number of suporters, but $D_Y > D_L$. According to [1], this corresponds to a typical situation in which Y wins most of times because $D_Y > D_L$, despite both initial parties had the same number of supporters. The total number n of realizations was always 1000 for every statisticallly studied issue trougout this work. The figure depicts the corresponding average time evolution of the parties sizes (A) and dimensions (D) considering just the cases in which either Y , L or G turned out to win. Specifically, the averages are computed with the formula:

$$\langle X \rangle = \frac{1}{n_K} \sum_{i=1}^{n_K} X_i \qquad K = Y, L, G \qquad (4)$$

where index i runs within the class of $n_K$ realizations in which party K finally turned out to win (note that $n_K \leq n$), and X is any given observable variable under consideration (i.e. D(t), A(t) of any particular party, at a given time t, etc.). For the cases in which Y finally wins, it initially shrinks mainly because of the arising of G. The same will in average happen for the cases in which L turns out to win. Whatever be the case, the winner party (L, Y or G) naturally needs to increase their population at the end. But Y is the party that wins most of times, and according to Fig.(3, right), it does it with an overall reduction of its dimensionality. We wonder if this corresponds with the general idea that many times, in order to win the election, a party incorporates as much new members as possible, loosing likely ideological purity. On Figs.(4) and (5), the above observation looks more general. It depicts overall average (Eq.(4)) variation of A and D as a function[6] of the mean initial dimensionality of Y and L[7]:

---

5 Specifically, a site (i, j) will belong to G if $M_{ij} = -M_{i'j'}$ or $M_{i'j'}$ , for every active neighbor (i′ , j ′) for which (i- i′ )² + (j- j′ )² is equal to 1 or 2, respectively.
6 Throughout this work, lines help to the reader to follow many numerically computed quantities. They do not represent model curves.
7 We consider $D_Y - D_L$ = const. = δD = 0.1 as in [1].

$$\bar{D} = \frac{D_L + D_Y}{2} \qquad (5)$$

within its correspondent possible range $J_r$ (Eq.(1)), for participation ratios r = 0.10, 0.15 and 0.20. Again, area increases for whatever the winning party be, but only Y always decreases its dimensionality.

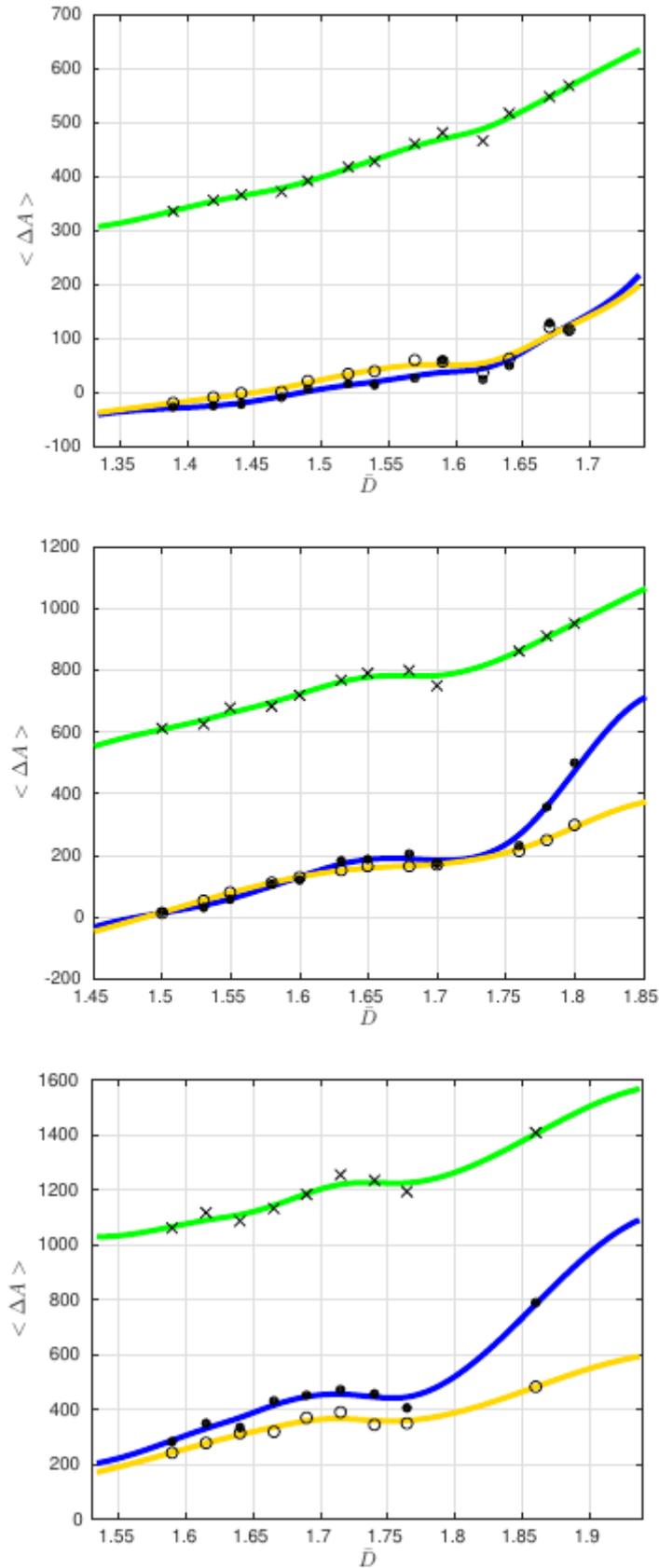

*Figure 4: Area variations (X = ΔA = $A_K$ (t = ∞) − $A_K$ (t = 0), K = Y, L, G in Eq.(4)), every time that either Y (∘), L (•) or G (×) party wins for three partipation ratios: r = 0.10 (top), r = 0.15 (middle) and r = 0.20 (bottom).*

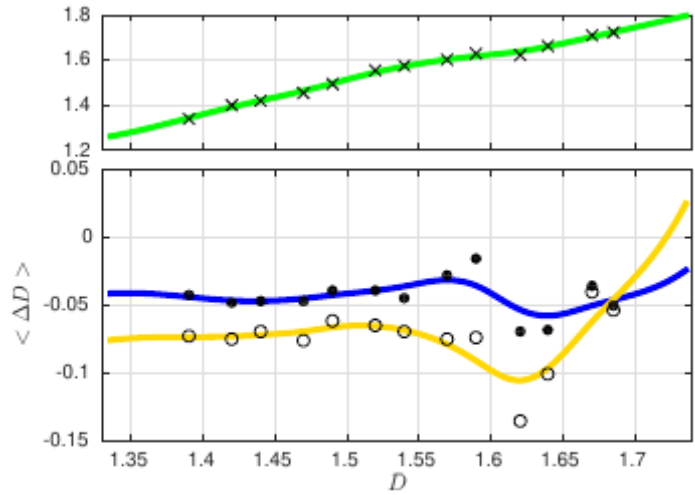

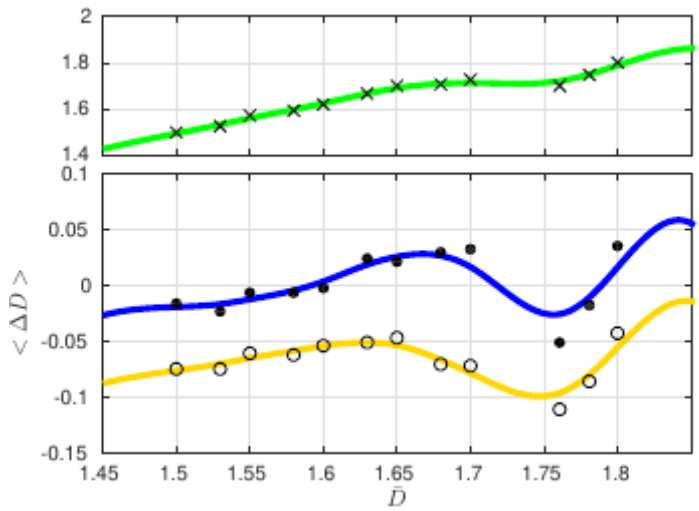

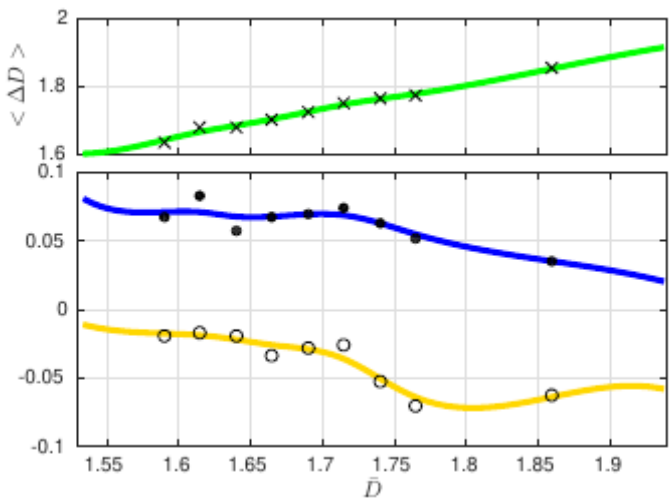

*Figure 5: Dimension variations ($X = \Delta D = D_K (t = \infty) - D_K (t = 0)$, $K = Y, L, G$ in Eq.(4)), every time that either Y (○), L (•) or G (×) party wins for three partipation ratios: $r = 0.10$ (top), $r = 0.15$ (middle) and $r = 0.20$ (bottom).*

Thus, Y seems to incorporate new members by relaxing at the same time its ideology because its dimensionality decreases when it wins. According to [1] we can recognize Y as the most sophisticated ideology party ($D_Y > D_L$), and it seems to be stronger than L for not only retaining its own less convinced supporters, but also for undergoing the welcome of people from other parties.

One may wonder why the political ideas commented just above fit very well between each other. There are not many parameters involved in the model at all. We claim that this may be a consequence of some general principles, rather than a particular dynamics shown by a particular system. In particular, we point out the existence of entropical reasons driving this issue.

*3.1. Entropy gradient*

We will focus on the ideas developed in [2], in which an entropical framework for describing a general class of complex systems is proposed. The reasons for this choice will look clearer below. By now, we start by saying that from that paper, it is possible to deduce (see Appendix A) a formula for the log number S(D, A) of all possible pattern configurations (of pixels) having fractal dimension D and area A:

$$S(D,A) = \frac{A}{U(D)} s(D) \qquad (6)$$

where U(D) and s(D) are positive functions defined in Eqs. (A.1) and (A.5) of Appendix A. Entropy S will not be associated to the total system. The formula will be applied to every particular party having size A and dimension D. Indeed, Eq.(6) will just be applied to the winner party.

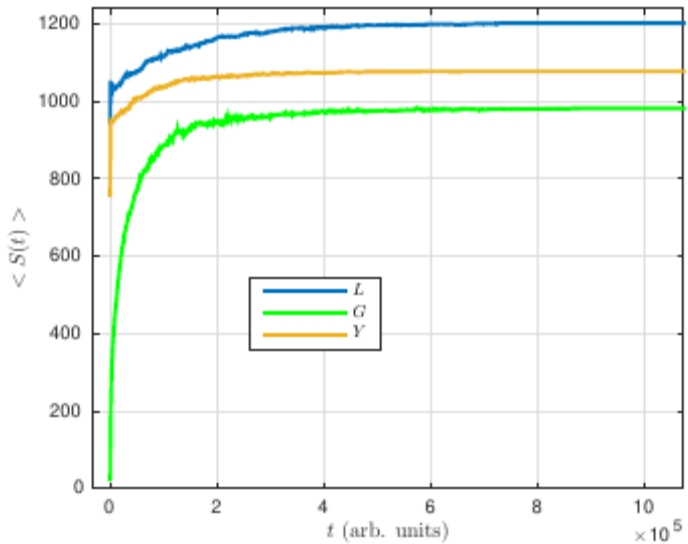

*Figure 6: Averaged (Eq.(4)) time dependence of entropy < S(t) > for the cases in which either Y, L or G turns out to win (r = 0.15, $D_Y$ = 1.7 and $D_L$ = 1.6, as in Fig.(3)).*

Fig.(6) was made by setting X = S(t) = S(D(t), A(t)) in Eq.(4). It shows that entropy increases in time for every winner party, assuming the same initial conditions of Fig.(3). In general, winner parties overall entropy variations will be always positive across their possible dimensional range $J_r$, independently of r. This can be observed on Fig.(7), which was made by taking X equal to the difference

$$\Delta S_K = S(D_K(t=\infty), A_K(t=\infty)) - S(D_K(t=0), A_K(t=0)) \qquad (7)$$

for K = Y, L, G in Eq.(4). One can see that every winner party evolves in time toward a more probable configuration, increasing its entropy analogously to a gas which irreversibly expands into a box.

Now, taking into account the above mentioned over Y and the fact that:

$$\frac{\partial S(D,A)}{\partial A} > 0 \qquad (8)$$

$$\frac{\partial S(D,A)}{\partial D} < 0 \qquad (9)$$

(see Fig.(11)), it is important to mention two points. First, the relative initial entropy values clearly determines the most probable winner (of the contend) between parties Y and L. In other words, Eq.(9) says that party Y is initially more organized (has a lower entropy) than L because $D_Y > D_L$. Secondly, Eqs. (8) and (9) say that the entropy gradient points out a natural evolution tendency:

mostly increasing A, which is an obvious requirement for winning for every party, and decreasing D, something that just Y party turns out to do it. L party does not work in the same way as Y's because its dimension variation is either positive or less negative than Y's (Fig.(5))[8].

In addition, assuming entropy gradient as one of the main ingredients for determining the track the most probably winner party will follow, we can see why Y entropy variation is higher than L's (e.g. Fig.(7), despite Y started from a lower entropy state, i.e.:

$$< \Delta S_Y > \quad > \quad < \Delta S_L > . \qquad (10)$$

We remark again this inequality is computed in the sense of Eq.(4). As a consequence, Y is more probable to win than L, reinforcing the gas thermodynamic analogy mentioned above.

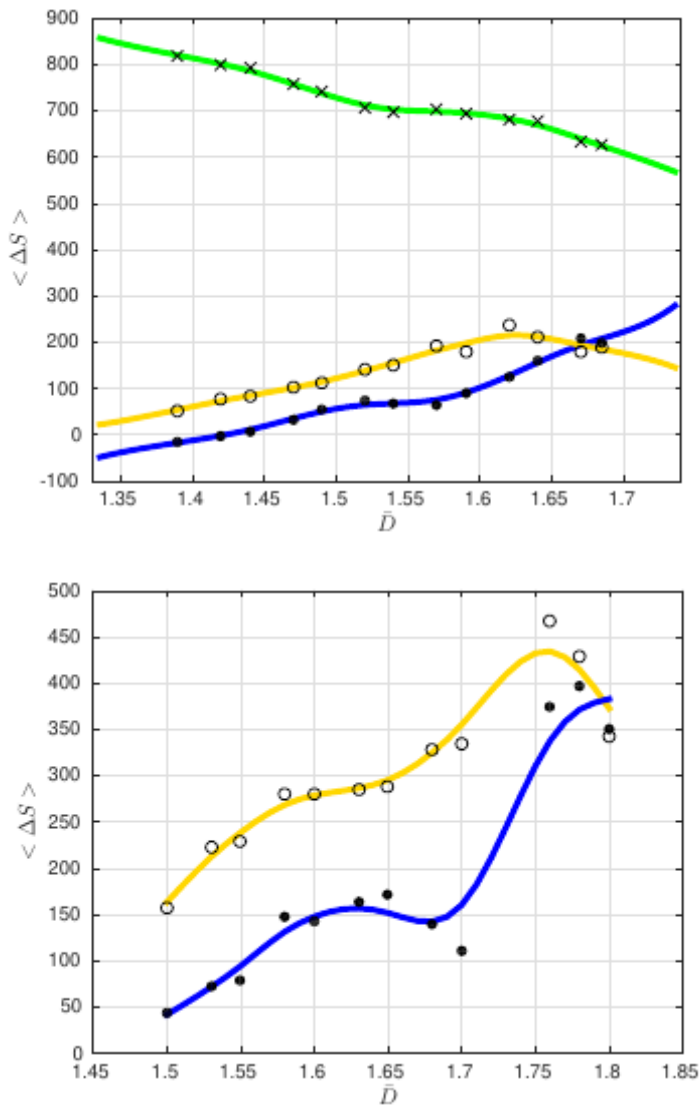

*Figure 7: Top: Entropy variations everytime that either Y (∘), L (•) or G (×) wins for r = 0.10. Bottom: Corresponding L's and Y's $< \Delta S >$'s for r = 0.15 (note the apparent non-monotonous behaviors with respect to D).*

*3.2. Winning probabilities of the parties and their relative entropy productions*

The probabilities of winning pK for every party (K = Y, L, G) are shown on Fig.(8). As it was studied in [1], these winning probabilities depend on $\bar{D}$, which in turn varies across a given Jr. We conveniently condense the probabilities per party by plotting them in terms of $\bar{D} - D_{min}$ (Eq.(1)).

---

8 There are just few cases in which size variation of Y is negative but L variation is more negative (Fig.(4, top)), so the entropy gradient still points out the winner track tendency.

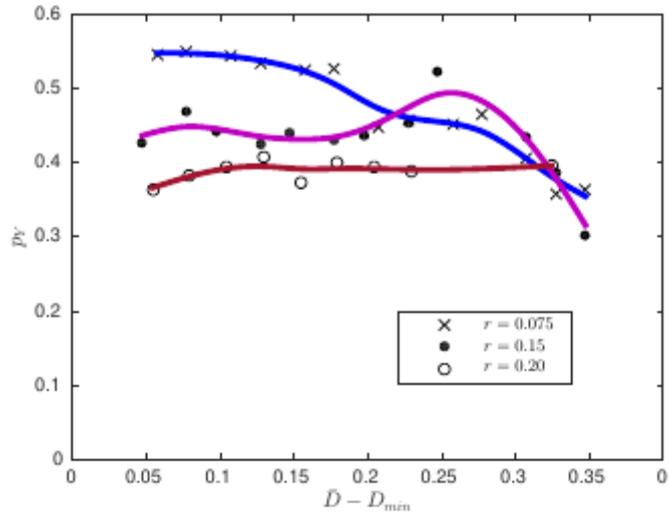

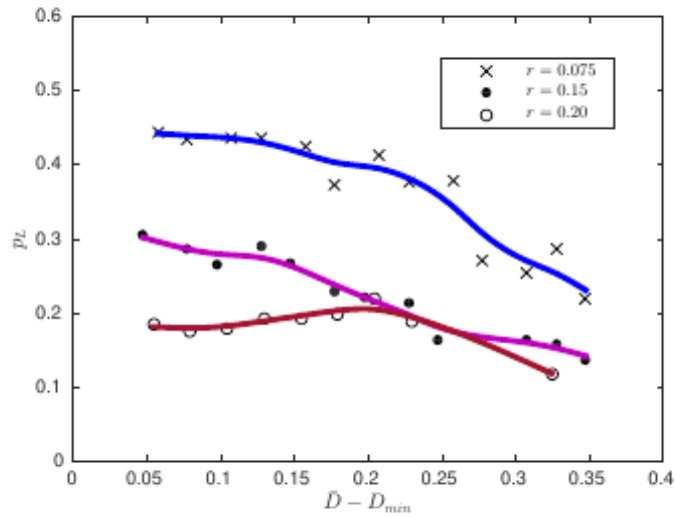

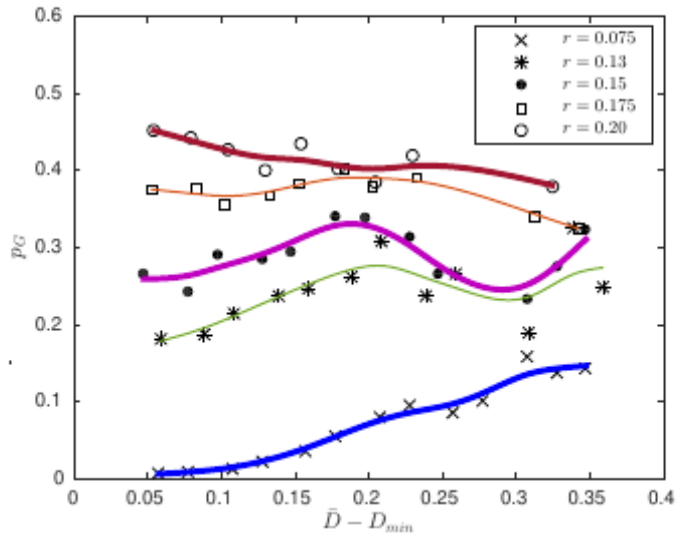

*Figure 8: The probability of winning of every party plotted with respect to $D_{min}$ for different participation ratios (r).*

The probabilities of winning of the parties show several types of regimes with respect to both dimensionality and participation ratio r (see [1]). As above, let us explain them in terms of entropy production. Actually, in order to explore the mechanisms that drive a given party to victory, we can delve into the meaning of Eq.(10) and study the relative entropy production between two any given possible winning parties (Y, L or G). We quantify this idea as

$$< \Delta S_{KK'} > = p_K < \Delta S_K > - p_{K'} < \Delta S_{K'} > \qquad K, K' = Y, L, G \qquad (11)$$

where unlike Eq.(10), every entropy production is now weighed according to the probabilities of winning of the parties.

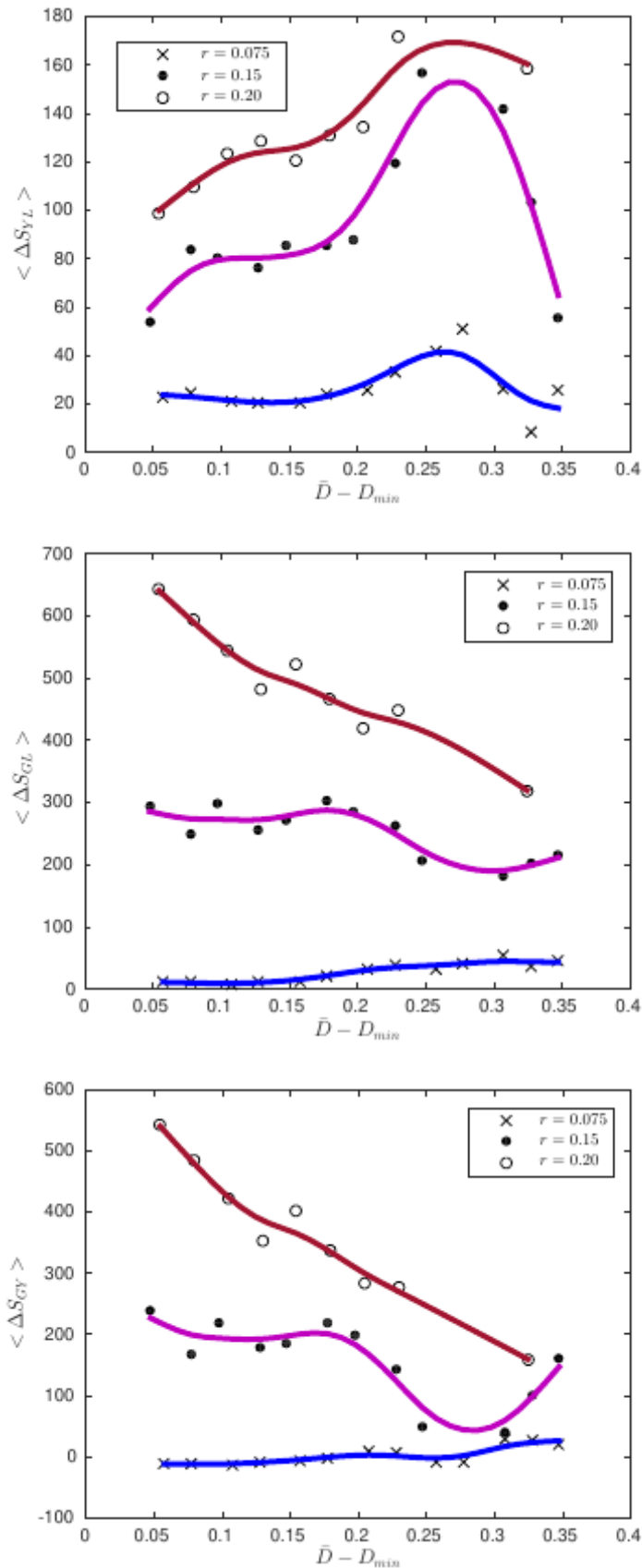

*Figure 9: Relative overall entropy variations (Eq.(11)) with respect to $D_{min}$ for different ratios (r).*

On Fig.(9) we show every possible $<\Delta S_{KK'}>$ for three partipation ratios r, as a function of $\bar{D} - D_{min}$. Note there is a general increasing behavior with respect to the participation ratio r, which clearly corresponds to the fact that S increases with A (see Eq.(6)).
Secondly, it is remarkable relative entropy productions of G look clearly different in shape and magnitude, reinforcing its emergent nature. Only initial party relative entropy productions show a strong dependence with the dimensionalities.

For extremely low participation ratios (× - curves of Fig.(8) and (9)), the dynamics is known to be stuck (see also [1]). This clearly corresponds to comparatively lower relative entropy productions between the parties. In other words, every pair of parties rearranges for winning, contributing to disorder in almost the same way. Consequently, system's dynamics shows a weak dependence on the initial dimensionalities of Y and L. This seems to explain why $p_Y$ and $p_L$ features look similar between each other. In addition, Fig.(9, middle, bottom) suggests that G has to rearrange comparatively more than Y or L, per unit of relative entropy production. This seems to explain why the emergent party G has a remarkably lower probability of winning than the rest.

In contrast, for medium low participation ratios (r = 0.15), the system shows a remarkable $\bar{D}$ - dependence (• - curves of Figs. (8) and (9)). There are substantially more dynamics and more probable emergence of new 3rd ideas ([1]), etc. that correspond to comparatively higher relative entropy productions. The existence of non-montonous behaviours on $J_{0.15}$ points out a kind of maximum in the myriad of possible initial configurations and subsequent system's evolution tracks.
It is possible to observe on Fig.(9) that every relative entropy variation is almost constant for low dimensionalities, while on the right part of $J_{0.15}$, they present a maximum ($\Delta S_{YL}$) or minima ($\Delta S_{GL}$ and $\Delta S_{GY}$).
Moreover, the increasing and decreasing regimes of • -curves of Figs.(9, top) and Fig.(8, top) clearly correspond between each other. The same happens between the • -curves of Fig.(9, middle, bottom) and Fig. (8, bottom).

Finally, for the highest low participation ratios[9] (r = 0.20), system's $\bar{D}$ - dependence decreases back (∘ - curves of Figs. (8) and (9)). Non-monotonicity features disappear. The approximately linear relative entropy productions produces an almost constant $\bar{D}$ - dependence for the winning probabilities. This means that the most important fact here is that $D_Y > D_L$ and not the value of $\bar{D}$. The latter also explains why $p_G$ becomes comparable to $p_Y$. Actually, despite the natural increments of relative
entropy productions, r = 0.20 lies at the limit of the high apathy regimes, for which the assumptions about random or hierarchically organized initial configurations become less relevant [1].

In summary, dynamics' $\bar{D}$ - dependence remarkably lessens at both extreme of the high apathy regime because of different already mentioned reasons. Only for intermediate partipation ratios, dimensionality dependence seems to produce a kind of optimal and diverse set of system outputs.

*3.3. Maximum imprevisibility regime*

The just above mentioned regimes clearly depends on the participation parameter r. This idea in principle sounds different from the regimes related to hierarchical behaviours that have been associated with D so far. In order to the study the role that r plays in system dynamics, we will associate an averaged dimension $D_r$ to each interval $J_r$ (Eq.(1)):

$$D_r = \frac{D_{min} + D_{max}}{2} \qquad (12)$$

The yellow curve of Fig.(10) shows that $D_r$ increases with r. The relation between $D_r$ and r is apparently non-linear as the ones of $D_{min}$ and $D_{max}$ with respect to r [1]. Thus, we can in some way see the participation parameter as a dimensionality too and viceversa. Let us see how further this supposition could take us.

As commented in the Introduction, the source of D- non-monotonicity underlying many aspects of [1] had not been analyzed in that paper. We proppose the non-monotonicity is just a counterpart of the fact that entropy function s(D) developed in [2] has a non-monotonous behavior with respect to D. Moreover, in [2] it is shown how the self-similarity of the succesive derivatives of s(D) with respect to D, provides a natural way to define different regimes regarding the notion of controllability of the (complex) system. In the context of [1] and the present work, we can associate the latter with the notion of predictability/previsibility about the final result of the election (or ideas contend) and/or the third party emergence and victory, etc.. Of course, due to the fact that the work [2] is too general and abstract, we cannot expect that all meanings of the regimes exactly fit to the present particular context. However, we claim there are many aspects of the opinion dynamic system considered in our work that clearly corresponds to those regimes. For instance, according to [2], it is possible to define a range

$$I_{21} = \{D : D_2 \leq D \leq D_1 \}, \quad (13)$$

which[10] can be associated with a kind of most uncontrolable regime in which not only the system's impredictability is higher but it is also increasing. For $\lambda = 64$, it turns out that $D_1 = 1.77$ and $D_2 = 1.51$.

---
9  Recall that as in [1], we always move within high apathy regimes.

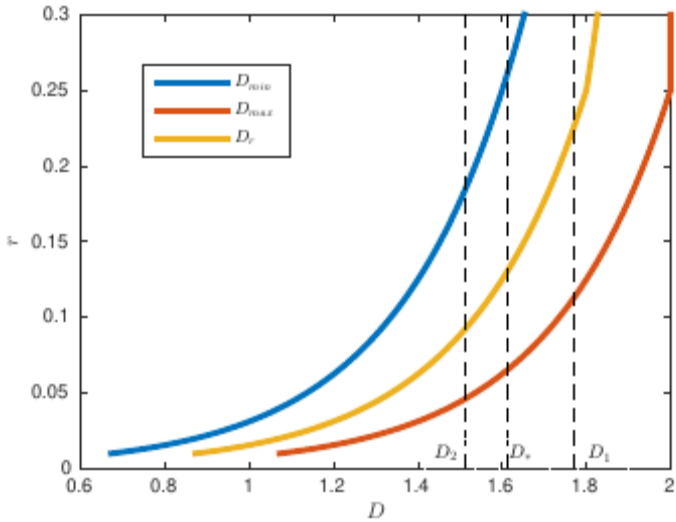

Figure 10: Relation between the participation degree r and dimension D. Note that this picture sums up the existence of an unpredictability regime $D_2 < D < D_1$.

Let us consider again the curve of $p_G$ for r = 0.15 shown on Fig.(8, bottom)[11] . The very fact that $(D_1 + D_2)/2 \approx 1.64$ is significant because it almost coincides with the local maximum of G probability of winning located at $\bar{D} \approx 1.63$ (it corresponds to $\bar{D} - D_{min} \approx 0.18$ on the figure). Thus, there is a local weakness associated to Y and L initial structures that seems to trigger the occurrence of a higher G probability of winning. In other words, the fact that $D_Y$ and $D_L$ initially lie in $I_{21}$ interval, makes the system to bear a kind of hidden indefinition that fosters an enhancement of G.

In addition we see from Figs. (8, bottom) and (10) that the non-monotonicity of $p_G$ is apparent when $D_r$ is clearly inside of the imprevisibility range $I_{21}$ (r = 0.15). If $D_r$ is closer to $D_1$ or $D_2$ or outside of $I_{21}$, $p_G$ will look more monotonous or constant (r = 0.075 and r = 0.20).

Figs. (7, bottom), (8) and (9) also show that non-monotonous behaviors arise when $D_r$ is inside of the imprevisibility range (r $\approx$ 0.15). All these non-monotonous features could be explained with the ideas of Section 3.1 as follows.

*3.4. A second order entropy effect*

The norm of the gradient of S(D, A) presents a minimum with respect to D that is almost independent of A (Fig.(11)). Remarkably, the minimum is located at $D = D_* \approx 1.61$, i.e almost at the center of the imprevisibility range. This fact clearly should have consequences on the entropy production of the winning party Y, which evolves in the direction of the gradient of S. Let us consider the r-dependent average

---

10  $D_1$ and $D_2$ are given by $s(D_1) = \max_D s(D)$ and $s'(D_2) = \max_D s'(D)$, respectively. There is an explicit formula for s(D) in Appendix A.

11  see also [1]

$$\langle \Delta S_K \rangle = \frac{\int_{D_{min}+\delta D/2}^{D_{max}-\delta D/2} \langle \Delta S_K \rangle d\bar{D}}{D_{max} - D_{min} - \delta D} \qquad K = Y, L, G, \qquad (14)$$

where the bracket inside of the integral is computed according to Eq.(4) varying $\bar{D}$ in a way that $D_L$ and $D_Y$ lie always within $J_r$ (Eq.(1)).

In correspondence to Section 3.2, Fig.(12, top) shows that every entropy variation has an increasing behavior with respect to r (or $D_r$ (Fig.(10)). The minimum gradient effect cannot be observed at glance. Actually, it is a 2nd order effect that can be unveiled after derivating with respect to r, as it is shown on Fig.(12, bottom): d $< \Delta S_K >_r /dr$ (K = Y, L) present either maxima (K = Y, L) or singular points (K = G) around r ≈ 0.14, i.e at the center of the imprevisibility range $I_{21}$ (Fig.(10)). However, note that the gradient acts primarily over the most structured initial party (Figs. (8, top) and (9, top)). To a less degree, the effect is also present in the emergent party which is a mixture of the initial parties (Figs. (8, bottom) and (9, center and bottom)). The gradient effects over the less structured initial party are weak and cannot be directly observed in terms of winning probability (Figs. (7, bottom)) and (8, center)).

Note that in correspondence to the already mentioned fact that $p_Y > p_L$, the quantities $< \Delta S_Y >_r$ and $< \Delta S_L >_r$ increase in different ways (Fig.(12, bottom)). Indeed for r ≥ 0.15, $< \Delta D_L >_r$ is positive, so L has by far less chances of being driven by the gradient than Y (Fig.(13)).

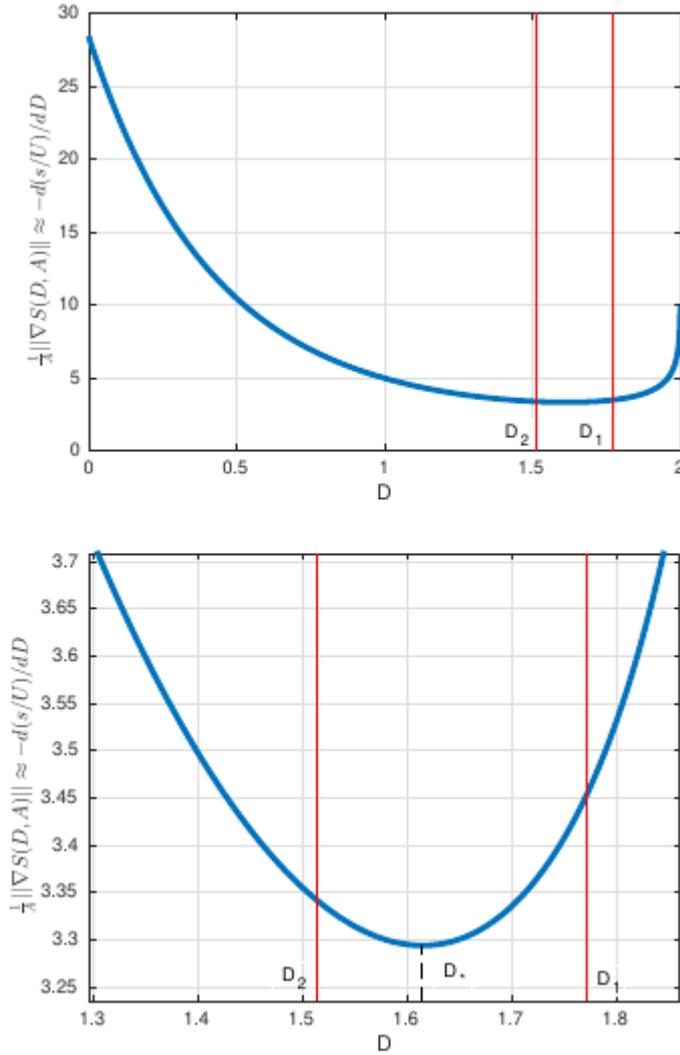

Figure 11: The norm of the gradient of S(D, A) has a minimum as a function of D that is almost independent of A (see Appendix B). Note these plots also show that $\partial S(D, A)/\partial D = A d(s/U)/dD < 0$ (Eq. (9)).

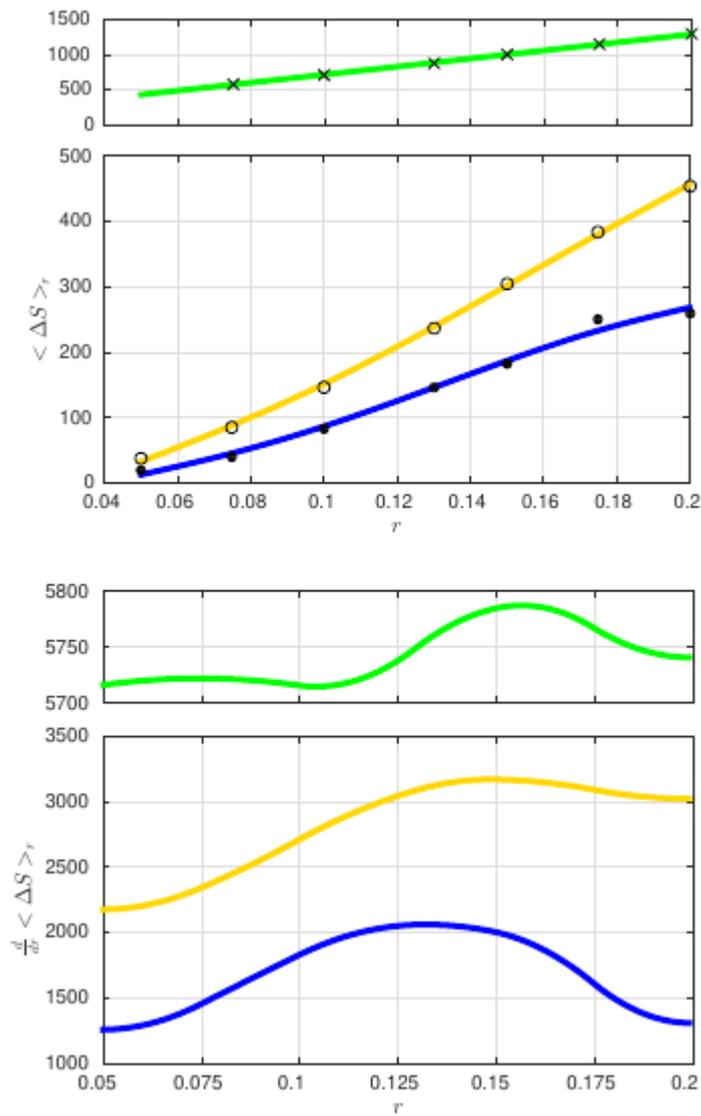

*Figure 12: $< \Delta S_K >_r$ and its derivative with respect to the participation ratio r, K = Y, L, G.*

## 4. Conclusions

The entropical analysis turned out to provide an explanation for not only the non-monotonous behaviors pointed out in [1], but for many others D-dependent regimes shown in the present work. The connexion between the winning probabilities of the parties and their relative entropy productions is paradigmatic (Figs. (8) and (9)).

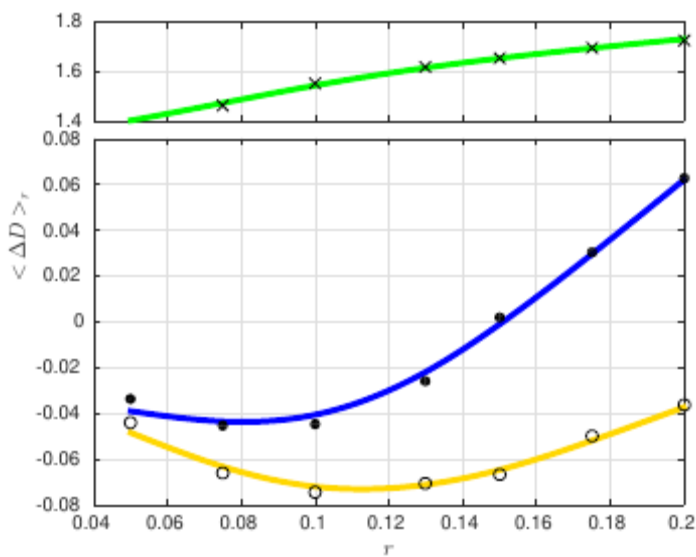

*Figure 13: r − dependent averages $< \Delta D >_r$ (as in Eq.(14)) for Y (∘), L (•) and G (×).*

The different entropical regimes point out an apparent track evolution dependence on the initial structure of the parties. This fact can clearly be associated to a kind of system's memory about both relative and absolute initial dimensionalities. Indeed, the most structured initial party is the one that wins most of the time, but the reasons seem to be more profound than the ones exposed in [1]. First, the higher initial dimension, the lower initial entropy. Secondly, the entropy gradient mostly drives the winning party divergent temporal evolution (see also Fig.(13)). Its positive relative entropy production with respect to L can be interpreted as a clear observable of that fact (Fig.(9, top)).

Conversely, as it was shown throughout this work, the less structured initial party has an associated narrow bunch of winning tracks that correspond both to less diverse victory set of possibilities and lower entropy productions. Even in the less probable cases in which L defeats Y by relaxing its ideology (r < 0.15 on Fig.(13)), it does it to a less degree than Y's. The correspondence to sociological situation looks clear. In order to gain supporters, more organized parties are always able to relax their ideals. In contrast, less organized positions necessarily have to modify their speeches according to the apathy level.

In turn, G party is different and inevitably has to organize for winning. This entropical analysis also enables to distinguish its emergent nature from the rest (Sections 3.2 and 3.3).

It was observed how the incontrollability ideas mentioned in [2] show up in the form of imprevisibility for present opinion dynamics model. The fact that the most diverse regimes in entropy production and winning probabibilities approximately lie within the range $D_2 \leq D \leq D_1$ is clearly significant, being in plenty agreement to [2].

For other interaction frameworks, the rising of incontrolability properties might be also plausible but just inside of $I_{21}$. It will be studied in [22].

We think that the moral of this study has to do with ubiquitous non-classical thermodynamical properties found in Complex Systems. These systems' hierarchical structures drive entropy production in anomalous ways. The assumptions of this work (and [1]) have a meaningful connexion with those hierarchical ideas.

**Acknowledgments**

*This work was supported by Consejo Nacional de Investigaciones Científicas y Técnicas (CONICET, Argentina). Authors thank Dr. H. S. Wio and Dra. D. A. Pedernera for stimulating and fruitful discussions.*

**Appendix A. Formula of S(D, A)**

According to [2], every pattern of A pixels and fractal dimension D lying in a λ × λ matrix, has to satisfy:

$$2^{mD} \leq A \leq U = U(D) = \lambda (2^m)^{D-2} ,  \quad (A.1)$$

where[12] $m = \log(\lambda/2)$. In addition, the number Ω of pixel configurations having dimensionality D and area A is given by Eq.(5) of [2]:

$$\Omega = \prod_{k=1}^{m} f\left(4, A 2^{-kD}, 2^D A 2^{-kD}\right) , \quad (A.2)$$

where f (x, y, z) turns out to be equal to the number of ways of setting z balls into y boxes with x compartments, leaving at least one ball per box. Eq.(8) of [2] shows that $f(x,y,z) \approx \log\left(\frac{xy}{z}\right)$ and that Stirling's approximation can be used (y and z z can be considered large numbers) to arrive at:

$$f(x,y,z) \approx xy\, H\left(\frac{z}{xy}\right) \quad (A.3)$$

---

[12] As in [2], log will be used as a shorthand for log 2.

where H is the Shannon's entropy function: $H(t) = -t \log t - (1-t) \log(1-t)$. Then, taking log in Eq.(A.2) it follows that

$$S(D,A) \approx \frac{A}{U} \sum_{k=1}^{m} \log f(4, U 2^{-kD}, 2^D U 2^{-kD}) \approx \frac{A}{U} s \qquad (A.4)$$

where $s = s(D)$ is the log number of pixels configurations having dimensionality D (see Eqs.(16-18) of [2]):

$$s(D) \approx \left(\frac{\lambda}{2^{m-1}}\right)^2 H(2^{D-2}) \frac{2^{mD}-1}{2^D-1} \qquad . \qquad (A.5)$$

**Appendix B. $\|\nabla S(D, A)\|$ approximation**

From Eqs. (A.1) and (A.4) one has that:

$\|\nabla S(D, A)\|^2 = A^2 (d(s/U)/dD)^2 + (s/U)^2 = (A/U)^2 [(s' - sm \ln 2)^2 + (s/A)^2]$. (B.1)

Now, s and s′ are shown to have the same order of magnitude in [2]. In turn, if A represents the initial size party (Y or L), one will have that $A^{-2} = (r\lambda)^{-2} << 1$ for every reasonable λ and r varying[13] in the high apathy regime [1]. Consequently, the right-side term envolving $A^{-2}$ of Eq.(B.1) can be neglected arriving at:

$\|\nabla S(D, A)\| \approx (A/U) |s - sm \ln 2| = A|d(s/U)/dD|$.

---

13  $(r\lambda)^{-2} \leq 0.05$ for $\lambda = 64$ and $0.075 \leq r \leq 0.2$